\documentclass[aps,twocolumn,superscriptaddress,superscriptemail,showpacs,showkeys]{revtex4}
\usepackage[dvipdf]{graphicx}
\usepackage{amssymb}
\usepackage{amsbsy}
\usepackage{latexsym}

\newcommand{\be}{\begin{equation}}
\newcommand{\ee}{\end{equation}}
\newcommand{\ba}{\begin{eqnarray}}
\newcommand{\ea}{\end{eqnarray}}

\begin{document}

\title{Chaotic universe in the $z=2$
Ho\v{r}ava-Lifshitz gravity}

\author{Yun Soo Myung}

\email{ysmyung@inje.ac.kr}

\affiliation {Institute of Basic Science and School of Computer
Aided Science, Inje University, Gimhae 621-749, Korea}

\author{Yong-Wan Kim}

\email{ywkim65@gmail.com}

\affiliation {Institute of Basic Science and School of Computer
Aided Science, Inje University, Gimhae 621-749, Korea}

\author{Woo-Sik Son}

\email{dawnmail@sogang.ac.kr}

\affiliation {Department of Physics and WCU-SSME Program Division, Sogang
University, Seoul 121-742, Korea}

\author{Young-Jai Park}

\email{yjpark@sogang.ac.kr}

\affiliation {Department of Physics and WCU-SSME Program Division, Sogang
University, Seoul 121-742, Korea}

\begin{abstract}
The deformed $z=2$ Ho\v{r}ava-Lifshitz gravity with coupling
constant $\omega$  leads to a nonrelativistic ``mixmaster"
cosmological model. The potential of  theory is given by the sum
of IR and UV potentials in the ADM Hamiltonian formalism. It turns
out that adding the UV-potential cannot  suppress chaotic
behaviors existing in the IR-potential.
\end{abstract}

\pacs{98.80.Qc, 98.80.Bp, 04.60.Pp, 98.80.Jk}

\keywords{Ho\v{r}ava-Lifshitz gravity, quantum gravity, mixmaster
universe}

\maketitle

\section{Introduction}

Recently, quantum gravity at a Lifshitz point, which is
power-counting renormalizable and hence potentially UV complete,
was proposed by Ho\v{r}ava~\cite{ho1,ho2,ho3}. This theory, unlike
string theory, is not intended to be a unified theory but quantum
gravity. Specific cosmological implications of $z=3$
Ho\v{r}ava-Lifshitz gravity with the Friedmann-Robertson-Walker
metric based on isotropy and homogeneity have recently been shown
in~\cite{cal,KK,muk}, including homogeneous vacuum solution with
chiral primordial gravitational waves~\cite{TS} and nonsingular
cosmological evolution with the big bang of standard and
inflationary universe replaced by a matter bounce~\cite{Bra,Rama,LS}. As far
as the cosmological solutions are concerned, there is no difference between
$z=2$~\cite{ho1} and $z=3$~\cite{ho2} Ho\v{r}ava-Lifshitz
gravities because the Cotton tensor vanishes when using the
Friedmann-Robertson-Walker metric. Furthermore, one has introduced
the deformed $z=3$ Ho\v{r}ava-Lifshitz gravity to find
asymptotically flat background~\cite{KS,Myungbh}.

On the other hand, the equations of general relativity lead to
singularities when we look at the equations backwards the origin of
time. Especially, we concentrate on a temporal singularity of the
solutions to the Einstein equations for the mixmaster model (Bianchi
IX Universe) describing an anisotropic and homogeneous cosmology. It
was well known that the approach to singularity shows a chaotic
behavior. The mixmaster
universe~\cite{mix1,mix2,mix3,mix4,cl,mix5,mix6,mix7} could be
described by a Hamiltonian dynamical system in a 6D phase space.
Belinsky, Khalatnikov, and Lifshitz (BKL) had conjectured that this
6D phase system could be well approximated by a 1D discrete Gauss
map that is known to be chaotic as one approaches the
singularity~\cite{BKL}. Chernoff and Barrow suggested that the
mixmaster 6D phase space could be split into the product of a 4D
phase space and a 2D phase space having regular
variables~\cite{mix3}. Following Cornish and Levin~\cite{cl}, Lehner
and Di Menza found that the chaos in the mixmaster universe is
obtained for the Hamiltonian system with potential having fixed
walls, which describes the curvature anisotropy~\cite{mix6}.
However, it turned out  that the mixmaster chaos could be
suppressed by (loop) quantum effects~\cite{BD,Bo}.

Hence it is very interesting to investigate cosmological application
of  Ho\v{r}ava-Lifshitz gravity in conjunction with  the mixmaster
universe based on the anisotropy and homogeneity because the
Ho\v{r}ava-Lifshitz gravity is a strong candidate for quantum
gravity.  However, we will not make any quantum operation on   the
Ho\v{r}ava-Lifshitz gravity.  In this work, we will make a progress
on this direction.

Getting an associated Hamiltonian within the ADM
formalism~\cite{adm} of deformed $z=2$ Ho\v{r}ava-Lifshitz
gravity~\cite{ho1,KS,Myungch}, we find two potentials in 6D phase
space: IR-potential $V_{IR}$ from 3D curvature $R$ and UV-potential
$V_{UV}$ from curvature square terms of $R^2$ and $R_{ij}R^{ij}$
with UV coupling parameter  $\omega$. In 4D phase space, we  find
 that the
UV-potential cannot suppress chaotic behaviors existing in the
IR-potential. After an extended analysis with movable wall, the
chaotic behaviors persist in the 6D phase space.

\section{Deformed $z=2$ Ho\v{r}ava-Lifshitz gravity}

The action of the deformed $z=2$  Ho\v{r}ava-Lifshitz
gravity~\cite{ho1,KS} takes the form in the (1+3)D spacetimes
\begin{eqnarray}
 S_\lambda &=& \int dtd^3x\sqrt{g}N\left[\frac{2}{\kappa^2}(K_{ij}K^{ij}-\lambda K^2)+\mu^3 R \right.\nonumber\\
           &&~~~~~ +\left.\frac{\kappa^2\mu^2(1-4\lambda)}{32(1-3\lambda)}R^2-\frac{\kappa^2\mu^2}{8}R_{ij}R^{ij}\right]
\end{eqnarray}
with three parameters $\kappa,~\mu,$ and $\lambda$. In the case of
$\lambda=1$, the above action leads to
\begin{eqnarray}\label{action}
 S_{\lambda=1}&=& \int  dtd^3x\sqrt{g}N\Bigg[\frac{2}{\kappa^2}(K_{ij}K^{ij}-K^2)\nonumber\\
              &&~~~~~~+\mu^3\left(R -\frac{2}{\omega}(R_{ij}R^{ij} -\frac{3}{8}R^2)  \right)\Bigg]
\end{eqnarray}
where the UV coupling  parameter $\omega=16\mu/\kappa^2$ is
introduced to control
 curvature square terms~\cite{Myungch}.
We note that $\omega$ is positive and thus, a negative $\omega$ is
not allowed for  the $z=2$ Ho\v{r}ava-Lifshitz gravity.
  In the limit of $\omega\to \infty(\kappa^2 \to 0)$,
 $S_{\lambda=1}$ reduces to general relativity (GR)  with the  speed of light $c^2=\kappa^2\mu^3/2$ and Newton's constant $G=\kappa^2/(32\pi c)$.
Also, we would like to  mention that the last line of (\ref{action})
seems to be similar to the action of 3D new massive gravity when
replacing $\mu^3$ and $\omega$ by $\frac{1}{16\pi c G}$ and $m^2$,
respectively~\cite{BHT}. However, although a similarity between them
exists, the difference is that in the $z=2$ Ho\v{r}ava-Lifshitz
gravity, the curvature $R$ is a nonrelativistic component
representing the 3D space in (1+3)D spacetimes, while in the 3D new
massive gravity, the 3D curvature $^3R$ is a relativistic one,
representing the (1+2)D spacetimes. Accordingly, the 3D new massive
gravity provides higher order temporal derivatives than second order
derivatives.

 Let us introduce the metric for the mixmaster universe  to
distinguish between expansion (volume change: $\alpha$) and
anisotropy (shape change: $\beta_{ij}$)
\begin{equation} \label{metric}
  ds^2=-dt^2+e^{2\alpha}e^{2\beta_{ij}}\sigma^i \otimes \sigma^j,
\end{equation}
where $\sigma^i$ are the 1 forms given by
\begin{eqnarray}
 && \sigma^1=\cos\psi d\theta+\sin\psi\sin\theta d\phi,\nonumber\\
 && \sigma^2=\sin\psi d\theta-\cos\psi\sin\theta d\phi,\nonumber\\
 && \sigma^3=d\psi+\cos\theta d\phi
\end{eqnarray}
on the three-sphere parameterized by Euler angles
($\psi,\theta,\phi$) with $0\leq\psi<4\pi$, $0\leq\theta<\pi$, and
$0\leq\phi<2\pi$. The shape change $\beta_{ij}$ is a $3\times 3$ traceless
symmetric tensor with det[$e^{2\beta_{ij}}]=1$ expressed in terms
of two independent shape parameters $\beta_\pm$ as
\begin{equation} \label{betapm}
 \beta_{11}=\beta_++\sqrt{3}\beta_-,~~\beta_{22}=\beta_+-\sqrt{3}\beta_-,~~\beta_{33}=-2\beta_+.
\end{equation}
The evolution of the universe is
described by giving $\beta_\pm$ as function of $\alpha$.
Note that the closed FRW universe is the special case of
$\beta_\pm=0$.

Now we concentrate on the behavior near singularity. Then, the
empty-space is sufficient to display the generic local evolution
close to a singularity  because the terms due to a matter or
radiation are negligible near singularity. Using Eq. (\ref{metric}),
the 3D curvature takes the form
\begin{equation}
 R= -12e^{-2\alpha}V_{IR}(\beta_+,\beta_-),
\end{equation}
where the IR-potential of curvature anisotropy is given by
\begin{eqnarray}
  V_{IR}(\beta_+,\beta_-) &=& \frac{1}{24}\Big[2e^{4\beta_+}\cosh(4\sqrt{3}\beta_-)+e^{-8\beta_+}\Big]\nonumber\\
                          &-& \frac{1}{12}\Big[2e^{-2\beta_+}\cosh(2\sqrt{3}\beta_-)+e^{4\beta_+}\Big].
\end{eqnarray}
The evolution of this universe is described by the motion of a point
$\beta=(\beta_+,\beta_-)$ as a function of $\alpha$ using the
time-dependent Lagrangian. The exponential wall picture of
IR-potential implies that a particle (the universe) runs through
almost free (Kasner) epochs where the potential could be neglected,
and it is reflected  at the walls, resulting infinite number of
oscillations. This  shows that the system under the IR-potential
behaves chaotically when the singularity is approached~\cite{dhn}.

Before we proceed, we briefly sketch  the IR-potential. We mention
that near the point $(\beta_+,\beta_-)=(0,0)$  corresponding to the
global minimum, the IR-potential takes an approximate form of
\begin{equation} \label{zzvir}
V_{IR}(0,0)\approx -\frac{1}{8}+(\beta^2_++\beta^2_-).
\end{equation}
 As is
shown in Fig. 1-(a), there are three canyon lines located at
$\beta_-=0$ and $\beta_-=\pm \sqrt{3}\beta_+$.  We discuss the
asymptotic structure of IR-potential.   In the case of $\beta_-\ll
1$, the IR-potential is either $V_{IR}\approx
2e^{4\beta_+}\beta^2_-$ if $\beta_+\rightarrow\infty$  or
$V_{IR}\approx \frac{1}{24}e^{-8\beta_+}$ if
$\beta_+\rightarrow-\infty$.  For $\beta_-=0$, one has
$V_{IR}\approx 0$ if $\beta_+\rightarrow\infty$. The potential is
bounded from below and exhibits discrete $Z_3$-symmetry by permuting
the principal axes of rotation $S^3$. Therefore, it has the shape of
an equilateral triangle in the anisotropy space ($\beta_+,\beta_-$)
and exponentially steep walls far away from the origin. A particle
 can only escape to infinity along the canyon lines
where the potential has the shape shown in Fig. 2 ($\omega=100$).
The smallest deviation from the axial symmetry will turn the
particle against the infinitely walls and thus, lead to a chaotic
motion. Another useful representation of the IR-potential is shown
in Fig. 3-(a) by drawing equipotential curves.  They extend
symmetrically between canyon lines at $\beta_-=0$ and $\beta_-=\pm
\sqrt{3}\beta_+$, which correspond  to a partially anisotropic
universe  with axial symmetry. The fully isotropic case is at the
origin (0,0), where the potential takes the global minimum.

The action (\ref{action}) provides  the time-dependent Lagrangian
\begin{eqnarray}
  {\cal L} &=& \mu^3e^{3\alpha}\left[-6(\dot{\alpha}^2-\dot{\beta}^2_+-\dot{\beta}^2_-)
           -12e^{-2\alpha}V_{IR}(\beta_+,\beta_-)          \right.\nonumber\\
           &&~~~~~~~~+ \left.\frac{e^{-4\alpha}}{16\omega}V_{UV}(\beta_+,\beta_-)\right],
\end{eqnarray}
where the dot denotes $\frac{d}{cdt}$.  One needs to introduce an
emergent speed of light $c$ in order to see the UV behaviors, while
for the IR behaviors, one chooses $c=1$ simply. Here, the
UV-potential of curvature square terms takes a complicated form
\begin{widetext}
\begin{eqnarray}
  V_{UV}(\beta_+,\beta_-) &=& \left[40 \Big(e^{8\beta_+}\cosh(4\sqrt{3}\beta_-)
                +e^{2\beta_+}\cosh(6\sqrt{3}\beta_-)+e^{-10\beta_+}\cosh(2\sqrt{3}\beta_-)\Big)
                -40e^{2\beta_+}\cosh(2\sqrt{3}\beta_-)\right.\nonumber\\
               && \left.+4e^{-4\beta_+}\cosh(4\sqrt{3}\beta_-)+2e^{8\beta_+}-20e^{-4\beta_+}
                   - 42e^{8\beta_+}\cosh(8\sqrt{3}\beta_-) -21e^{-16\beta_+}\right],
\end{eqnarray}
\end{widetext}
which is a key feature of the deformed $z=2$ Ho\v{r}ava-Lifshitz
gravity. Note that considering the deformed $z=3$
Ho\v{r}ava-Lifshitz gravity~\cite{KS}, one would expect to have a
more complicated UV potential because of the presence of the Cotton
tensor~\cite{BBLP}. However, we have shown that the Cotton tensor
does not change  significantly the situations~\cite{MKSP}.

 In order to
appreciate implications of chaotic approach to the deformed $z=2$
Ho\v{r}ava-Lifshitz gravity, we have to calculate the Hamiltonian
density by introducing three canonical momenta as
\begin{equation}
  p_\pm=\frac{\partial {\cal L}}{\partial\dot{\beta}_\pm}=12\mu^3e^{3\alpha}\dot{\beta}_\pm,
  ~~~p_\alpha=\frac{\partial {\cal
  L}}{\partial\dot{\alpha}}=-12\mu^3e^{3\alpha}\dot{\alpha}.
\end{equation}
The normalized canonical Hamiltonian in 6D phase space takes the
form
\begin{equation}
  \label{6DHam}
  {\cal H}_{6D}
  =\frac{1}{2}(p^2_++p^2_--p^2_\alpha)+V_\alpha(\beta_+,\beta_-,\omega), \end{equation}
  where the potential is given by
\begin{equation} \label{6DHamp}
    V_\alpha(\beta_+,\beta_-,\omega)=e^{4\alpha}\Big(V_{IR}-\frac{e^{-2\alpha}}{192
    \omega}V_{UV}\Big).
\end{equation}
Here ${\cal H}_{6D}=12\mu^3e^{3\alpha}{\cal H}_c$ using the
canonical Hamiltonian ${\cal H}_c$.  In this case, let us choose the
parameter $12\mu^3=1$ for simplicity.  Then, the Hamiltonian
equations of motion are
\begin{eqnarray}
 \label{eom1}
 && \dot{\beta}_\pm=p_\pm,~~\dot{p}_\pm=-e^{4\alpha}\frac{\partial V_{IR}}{\partial\beta_\pm}
  +\frac{e^{2\alpha}}{192\omega}\frac{\partial V_{UV}}{\partial\beta_\pm},\nonumber\\
 && \dot{\alpha}=-p_\alpha,~~\dot{p}_\alpha=-4e^{4\alpha}V_{IR}
                   +\frac{e^{2\alpha}}{96\omega}V_{UV}
\end{eqnarray}
in 6D phase space.

\begin{figure}[t]
\includegraphics{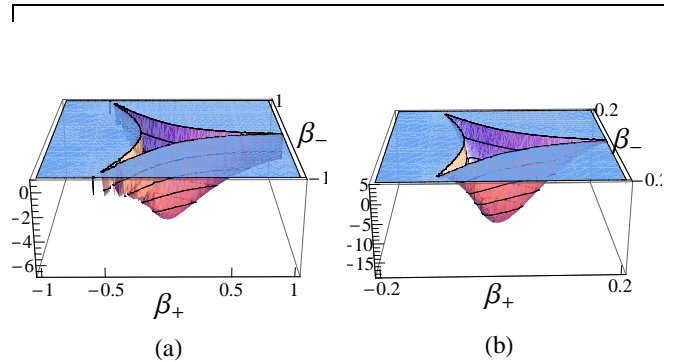}
\caption{3D Potential shapes $V(\beta_+,\beta_-,\omega)$ with three
canyon lines located at $\beta_-=0$ and $\beta_-=\pm
\sqrt{3}\beta_+$: (a) for $\omega=100$, it corresponds to the GR
mixmaster model ($V_{IR}$ dominates). (b) For $\omega=0.01$, it
corresponds to the potential of $z=2$ Ho\v{r}ava-Lifshitz gravity.}
\label{fig:2Dpotential.eps}
\end{figure}

\section{Chaotic behaviors in 4D phase space}

 Chernoff and Barrow showed that the mixmaster 6D phase
space could be split into the product of a
 4D phase space showing chaotic behavior and a 2D phase space showing regular
behavior~\cite{mix3}. Hence, we confine  the dynamical system to a
4D phase space. Setting $\alpha = 1$, we consider the motion of a
particle (the universe) of coordinates ($\beta_+,\beta_-$) under the
full potential of
\begin{equation}
V_{\alpha=1}(\beta_+,\beta_-,\omega) \to
V(\beta_+,\beta_-,\omega)=e^4\Bigg[V_{IR}-\frac{V_{UV}}{192 e^2
\omega}\Bigg].
\end{equation}

The key to chaos in mixmaster dynamics is that the potential has
been developed ``corners", that is, places where very special
trajectories do not encounter walls. This leads to sensitive
dependence on initial conditions, because a tiny change in how
closely a trajectory approaches a corner can lead to large changes
in the sequence of bounces off the walls. On the other hand, the
behavior of the potential near the origin has little to do with
chaos because the overall decrease in scale leads to the walls and
the trajectories are moving ever farther outward from the origin.

Let us describe the $z=2$ Ho\v{r}ava-Lifshitz potential
$V(\beta_+,\beta_-,\omega)$ intensively.  In order to make a
connection to the loop quantum gravity,  we first mention that near
the origin $(\beta_+,\beta_-)=(0,0)$ which corresponds to the closed
FRW universe, the potential $V(\beta_+,\beta_-,\omega)$ takes
approximately the form of
\begin{equation}
\label{zzv}
 V(0,0,\omega)\approx
-\Big(\frac{e^{4}}{8}+\frac{e^2}{64\omega}\Big)
+\Big(e^4+\frac{17e^2}{4\omega}\Big)\Big(\beta^2_++\beta^2_-\Big).
\end{equation}
Comparing (\ref{zzv}) with (\ref{zzvir}), the former reduces to the
latter up to $e^4$ in the limit of $ \omega \to \infty$. It turns
out that adding the UV-potential makes just the potential well at
the origin deeper, compared to the IR case.  At this stage, it is
curious to ask whether the inflection point at the origin of
$(\beta_+,\beta_-)=(0,0)$ exists, which might show a change from
chaotic behavior to non-chaotic behavior. This point may be
determined by the condition of
\begin{equation}
V''(\beta_+,0,\omega)|_{\beta_+=0}=V''(0,\beta_-,\omega)|_{\beta_-=0}=0,
\end{equation}
which leads to
\begin{equation}
2e^4+\frac{17e^2}{2\omega}=0.
\end{equation}
However,  we have the negative $\omega_c$
\begin{equation}
\omega_c=-\frac{17}{4e^2} \simeq -0.5752.
\end{equation}
This may imply  that there is no inflection point which makes a
transition from chaotic behavior to non-chaotic behavior.

\begin{figure}[b]
\includegraphics{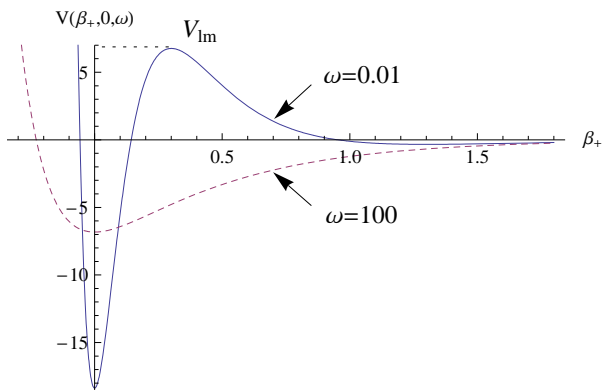}
\caption{Potential graphs $V(\beta_+,\beta_-,\omega)$ with
$\beta_-=0$: the long-dashed curve is for $\omega=100$ (GR) and the
solid curve for $\omega=0.01$ with a local maximum $V=V_{lm}$.}
\label{fig:potential.eps}
\end{figure}
\begin{figure}[t]
\includegraphics{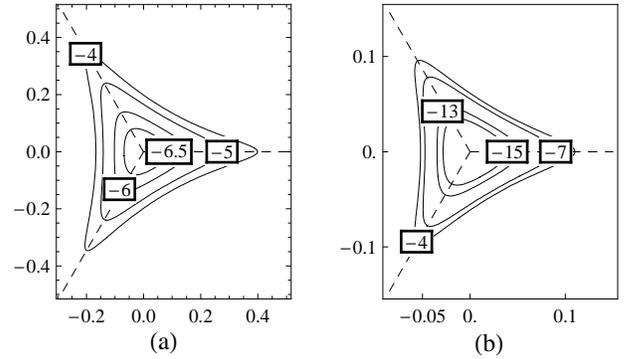}
\caption{Equipotential curves, developing triangles with corner: (a)
for $\omega=100$ with $E=-6.5,-6.0,-5.0,-4.0$, respectively, (b) for
$\omega=0.01$ with $E=-15.0,-13.0,-7.0,-4.0$, respectively. Three
canyon lines are developed along corners at $\beta_-=0$ and
$\beta_-=\pm \sqrt{3}\beta_+$.} \label{fig:contourAll.eps}
\end{figure}
\begin{figure}[t]
\includegraphics{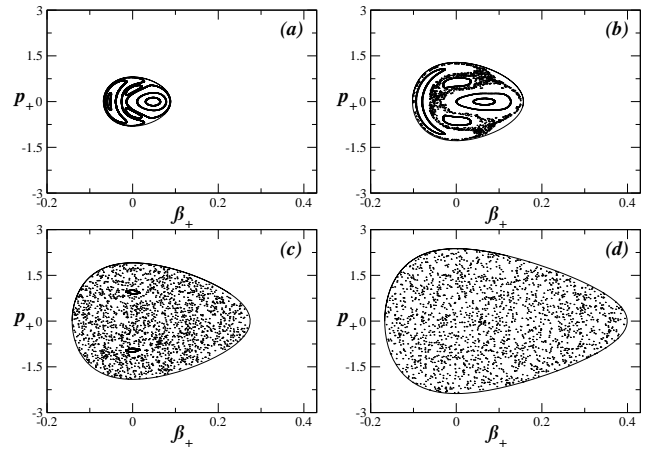}
\caption{Poincar\'{e} sections for the $\omega=100$ case with (a) $E=-6.5$,
(b) $E=-6.0$, (c) $E=-5.0$, and (d) $E=-4.0$.} \label{fig:Poin_100.eps}
\end{figure}
  The asymptotic structure
of the full  potential is given  as follows. For $\beta_-=0$, if
$\beta_+\rightarrow\infty$, then \be
V(\beta_+,\beta_-,\omega)\approx 0. \ee For $\beta_-\ll 1$, if
$\beta_+\rightarrow\infty$,  one finds
\begin{equation}
 e^{-2}V(\beta_+,\beta_-,\omega)\approx
 \frac{16\beta^2_-}{\omega} e^{8\beta_+}.
\end{equation}
For $\beta_-\ll 1$, if $\beta_+\rightarrow -\infty$, then
\begin{equation}
 e^{-2}V(\beta_+,\beta_-,\omega)\approx
     \frac{7}{64\omega}e^{-16\beta_+}.
\end{equation}
Therefore, the asymptotic structure is determined by the
UV-potential.
 As is shown in Fig. 1-(b),
there are three canyon lines located at $\beta_-=0$ and $\beta_-=\pm
\sqrt{3}\beta_+$.
 The potential is bounded from below and exhibits
discrete $Z_3$-symmetry by permuting the principal axes of rotation
$S^3$. Therefore, it has the shape of an equilateral triangle in the
anisotropy space ($\beta_+,\beta_-$) and exponentially steep walls
far away from the origin. However,  for total energy $E<V_{lm}$, a
particle  cannot
 escape to infinity along the canyon lines where the potential has
a local maximum $V=V_{lm}$ as a bump shown in Fig. 2
($\omega=0.01$).  For $E<0$, the smallest deviation from the axial
symmetry will turn the particle against the infinitely walls and
thus, lead to a chaotic motion. Although a local maximum appears
along canyon lines, equipotential curves in Fig. 3-(b) are similar
to Fig. 3-(a) of IR-potential for $E<0$. Hence, we expect that
chaotic behavior appears in the $z=2$ Ho\v{r}ava-Lifshitz gravity
with small $\omega$. The dynamics of particle seems complicated for
$0<E<V_{lm}$ and thus, we skip it.

In general, the chaos could be defined as being such that (i) the
periodic points of the flow associated to the Hamiltonian are dense,
(ii) there is a transitive orbit in the dynamical system, and (iii)
there is sensitive dependence on the initial condition.
 Our reduced
system is described by the 4D Hamiltonian
\begin{equation}
  {\cal H}_{4D} =\frac{1}{2}(p^2_++p^2_-) + V(\beta_+,\beta_-,\omega).
\end{equation}
It is well known  that the appearance of  chaotic behavior in the
mixmaster dynamics is closely related to the appearance of ``corner"
potential.  To make a definite connection, we choose $\omega=100$
and $\omega=0.01$ for the original mixmaster universe of GR  and the
mixmaster universe of  $z=2$ Ho\v{r}ava-Lifshitz gravity,
respectively.  As is shown in Fig. 3, when making equipotential
curves,  we find that the corner appears for $E=-5.0$ and $-4.0$ for
$\omega=100$ and  for $E=-7.0$ and $-4.0$ for $\omega=0.01$, which
implies the appearance of chaos.

Let us perform simulations of the 4D dynamics and represent
Poincar\'{e} sections, which describe the trajectories in phase
space $(p_+,\beta_+)$ by varying the total energy $E$ or ${\cal
H}_{4D}$ of the system.  Actually, we have performed the analysis
for the $\omega=100,1,0.03,$ and $0.01$ cases. We have found that
the chaotic behavior persists for all $\omega>0$. Figs. 4 and 5
present two typical cases, showing that the intersections of several
computed trajectories are  displaced in ($p_+,\beta_+$) with the
plane $\beta_-=0$ for different values of energies. In each plot, we
choose initial points which correspond to a prescribed kinetic
energy. Also we confirm that for $\omega=0.01$,  complicated chaotic
behaviors appear for $0<E< V_{lm}$.

The results of Poincar\'{e} sections show that considering  lower
energies $E=-6.5$ and $-6.0$ for $\omega=100$ and $E=-15.0$ and
$-13.0$ for $\omega=0.01$ within the potential well, the integrable
behavior dominates and the intersections of trajectories represent
closed curves.  Importantly, concerning higher energies $E=-5.0$ and
$-4.0$ for $\omega=100$ and $E=-7.0$ and $-4.0$ for $\omega=0.01$
within the potential well, {\it the closed curves are broken up
gradually and the bounded phase space fills with a chaotic sea}. The
same kinds of plots have been obtained for the other phase space
($p_-,\beta_-$) with the plane $\beta_+=0$. At this stage, we
mention that for $\omega=100$, the case of $E=0$ leads to the
``corners", where special trajectories do not encounter walls.
However, for $\omega=0.01$, the case of $E=0$ does not lead to the
``corner" because of the presence of a local maximum. For this, see
Fig. 2.

Finally,  it confirms that the appearance of chaotic behavior in the
mixmaster dynamics of $z=2$ Ho\v{r}ava-Lifshitz gravity  is closely
related to the appearance of corner.
\begin{figure}
\includegraphics{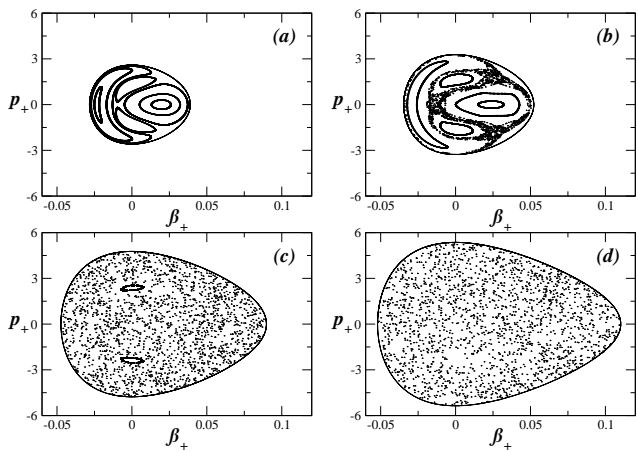}
\caption{Poincar\'{e} sections for $\omega=0.01$ with  (a)
$E=-15.0$, (b) $E=-13.0$, (c)
$E=-7.0$, and (d) $E=-4.0$.}
\label{fig:Poin_0.01.eps}
\end{figure}

\section{Chaotic behavior in 6D phase space}
Up to now, we have investigated  the dynamics at fixed $\alpha=1$
for simplicity. This means that we have never explored the outer
region of the potential which also determines whether the $z=2$
Ho\v{r}ava-Lifshitz gravity is chaotic. We remind the reader that
the true phase space is 6D for the vacuum universe, and thus, we
have to consider a movable billiard with the potential
$V_\alpha(\beta_+,\beta_-,\omega)$ in Eq. (\ref{6DHamp}) because
the walls are moving with time  since the logarithm of the volume
change $\alpha=\frac{1}{3}\ln V$ and its derivative are entering
in the system.  In this case, $\alpha$ and $p_\alpha$ are regular
variables as functions of time. Therefore, one should perform a
full simulation which includes the dynamics of $\alpha$ and
$p_\alpha$. However, this seems to be a formidable task and thus,
we could not make a progress on this direction.

In this section, instead,  we will investigate  a possibility of
finding chaotic behaviors by considering  the small volume limit of
$\alpha \to -\infty$ only. To this end, it would be better to
introduce a new time $\tau$ defined by
\begin{equation}
\label{newtime} \tau=\int \frac{dt}{V},~~V=e^{3\alpha},
\end{equation} which makes decoupling of the
volume $\alpha$ from the shape $\beta_\pm$  explicitly. Starting
from the action (\ref{action}) and integrating out the space
variables, we have
\begin{eqnarray}
\label{actionB}
  \bar{S}_{\lambda=1} &=& \mu^3 \int d\tau
                  \frac{e^{3\alpha}N}{V}
                 \left[6(-\alpha^{'2}+\beta^{'2}_++\beta^{'2}_-)
                 \right. \nonumber\\
                    &-& \left. V^2\left(12e^{-2\alpha}V_{IR}(\beta_+,\beta_-)
                    - \frac{e^{-4\alpha}}{16\omega}V_{UV}(\beta_+,\beta_-)  \right) \right],
                    \nonumber\\
\end{eqnarray}
where the prime ($'$) denotes the derivatives with respect to
$\tau$. Plugging $N=1$ into (\ref{actionB}), we have the
Lagrangian as
\begin{eqnarray}
\label{Lagtau}
  \bar{\cal L}_{\lambda=1} &=& \mu^3
  \left[6(-\alpha^{'2}+\beta^{'2}_++\beta^{'2}_-)\right.  \nonumber\\
                    &-& \left. 12e^{4\alpha}\left(V_{IR}(\beta_+,\beta_-)
                    - \frac{e^{-2\alpha}}{192\omega}V_{UV}(\beta_+,\beta_-)
                    \right) \right]. \nonumber\\
\end{eqnarray}
The canonical momenta are given  by
\begin{eqnarray}
  \bar{p}_\pm = \frac{\partial \bar{\cal L}_{\lambda=1}}{\partial\beta'_\pm}
            =12\mu^3\beta'_\pm,~
  \bar{p}_\alpha = \frac{\partial \bar{\cal L}_{\lambda=1}}{\partial\alpha'}
            =-12\mu^3\alpha'.
\end{eqnarray}
Then, the canonical Hamiltonian in 6D phase space is obtained to
be
\begin{eqnarray}
  \bar{\cal H}_{6D}
    &=& \bar{p}_\alpha\alpha'+\bar{p}_+\beta'_++\bar{p}_-\beta'_--\bar{\cal L}_{\lambda=1}\nonumber\\
        &=&   \frac{1}{2}(\bar{p}^2_++\bar{p}^2_--\bar{p}^2_\alpha)
             +e^{4\alpha}\Big(V_{IR}-\frac{e^{-2\alpha}}{192 \omega}V_{UV}
             \Big),\nonumber\\
\end{eqnarray}
where we have chosen the parameter $12\mu^3=1$ for simplicity.
Then, the Hamiltonian  equations of motion are obtained as
\begin{eqnarray}
 &&\label{appdix1}
 \beta'_\pm=\bar{p}_\pm,
 ~~\bar{p}'_\pm=-e^{4\alpha}\frac{\partial V_{IR}}{\partial\beta_\pm}
               +\frac{e^{2\alpha}}{192\omega}\frac{\partial V_{UV}}{\partial\beta_\pm},\nonumber\\
 &&
 \alpha'=-\bar{p}_\alpha,
 ~~\bar{p}'_\alpha =-4e^{4\alpha}V_{IR} +\frac{e^{2\alpha}}{96\omega}V_{UV}.
\end{eqnarray}
We note that comparing Eqs. (\ref{appdix1}) with Eqs. (\ref{eom1}),
there is no change in the Hamiltonian and its equations of motion
except replacing $t$ by $\tau$.  The evolution of $\alpha$ is in
particular determined by
\begin{equation}
\alpha''=4e^{4\alpha}V_{IR}
                   -\frac{e^{2\alpha}}{96\omega}V_{UV}.
\end{equation}
Then, we obtain a 6D phase space consisting in the product of a 4D
chaotic one times a 2D regular phase space for the $\alpha$ and
$p_\alpha$ variables.  As the volume goes to zero near singularity
($e^{4\alpha}\to 0,~p_\alpha \to 0$),  one finds the limit
\begin{equation}
\bar{{\cal H}}_{6D} \to
\frac{1}{2}\Big(\bar{p}_+^2+\bar{p}_-^2\Big)+K \not={\cal H}_{4D}.
\end{equation}
Hence, we note that the 6D system is not asymptotic in  $\tau$ to
the previous 4D system.

Now, we are in a position to  show that the presence of the
UV-potential does not suppress chaotic behaviors existing in the
IR-potential.  For this purpose, we have to introduce two
velocities: particle velocity $v_p$ and wall velocity $v_w$ defined
by
\begin{equation}
v_p=\sqrt{\bar{p}_+^2+\bar{p}_-^2},~~v_w=\frac{d\beta_+^w}{d\tau},
\end{equation} where the wall location $\beta_+^w$ is determined by
the fact that the asymptotic potential $K$ is significantly felt
by the particle as
\begin{equation}
\label{asymK}
 \bar{p}^2_\alpha\approx 2K=
 \frac{e^{4\alpha-8\beta_+}}{12}+\frac{7e^{2\alpha-16\beta_+}}{32\omega}
\end{equation}
in the limit of $\beta_+ \to -\infty$. On the other hand, the
particle velocity is given by
\begin{equation} \label{particlev}
v_p=\sqrt{2\bar{{\cal H}}_{6D}+\bar{p}_\alpha^2-2K}.
\end{equation}
In the IR-limit ($\omega \to \infty$)
of Einstein gravity, the wall location is determined by
\begin{equation}
 \beta_+^w \approx \frac{\alpha}{2}-\frac{1}{8}\ln\Big[12\bar{p}_\alpha^2\Big].
\end{equation}
Then, the wall velocity is given by
\begin{equation}
 v_w^{IR}=-\frac{d\beta_+^w}{d\tau} \approx  \frac{\bar{p}_\alpha}{2}+\frac{e^{4\alpha-8\beta_+}}{24\bar{p}_\alpha},
\end{equation}
which leads to
\begin{equation}
|v_w^{IR}| \approx \frac{|\bar{p}_\alpha|}{2}.
\end{equation}
As a result, we find that the particle velocity is always greater
than the wall velocity as
\begin{equation}
v_p^{IR}=\sqrt{2{\bar{{\cal
H}}_{6D}+\bar{p}_\alpha^2-2e^{4\alpha}V_{IR}}} \approx
|\bar{p}_{\alpha}|>v_w^{IR}.
\end{equation}
Thus, there will be an infinite number of collisions of the
particle against the wall since it will always catch a
wall~\cite{mix6,mix7}.

Next, let us investigate what happens in the UV-limit of $\omega
\to 0$. In this case, the $V_{UV}$ term dominates. The wall
velocity takes the form \be |v^{UV}_w|=
\frac{|\bar{p}_\alpha|}{8}, \ee and the particle velocity leads to
\be v^{UV}_p\approx |\bar{p}_\alpha|>|v_w^{UV}| \ee in the limit
of $\alpha \to -\infty$. This case is similar to the Einstein
gravity.

Finally, we could not observe a slowing down of the particle
velocity due to the UV effects. This means that the chaos persists
in the moving wall.

\section{Discussions}
First of all, we point out that for $\omega>0$, there always exists
chaotic behavior.  This contrasts  to the case of the loop mixmaster
dynamics based on loop quantum cosmology~\cite{Bo}, where the
mixmaster chaos could be  suppressed by loop quantum
effects~\cite{BD}. In the loop quantum cosmology, the effective
potential at decreasing volume labeled by ``discreteness $j$"  are
significantly changed in the vicinity of (0,0)-isotropy  point in
the anisotropy plane $-\beta_+$. The potential at larger volumes
exhibits a potential wall of finite height and finite extension like
Fig. 2. As the volume is decreased, the wall moves inward and its
height decreases. Progressively, the wall disappears completely
making the potential negative everywhere at a dimensionless volume
of $(2.172j)^{3/2}$ in the Planck units. Eventually, the potential
approach zero from below. This shows that classical reflections will
stop after a finite amount of time, implying that classical
arguments about chaos inapplicable. Once quantum effects are taken
into account, the reflections stop just when the volume of a given
patch is about the size of Planck volume.

To that end,  the role of UV coupling parameter $\omega$ is
 different from the area quantum number $j$ of the loop
quantum gravity. In our case, time variable (related to the volume
of $V=e^{3\alpha}$) as well as two physical degrees of anisotropy
$\beta_{\pm}$ are treated in the classical  way without
quantization. However, in the loop quantum framework, all three
scale factors were quantized using the loop techniques. Hence two
are quite different: the  potential wells at the origin did not
disappear for any $\omega>0$ in the $z=2$ Ho\v{r}ava-Lifshitz
gravity, while in the loop quantum gravity the height of potential
wall rapidly decreases until they disappears completely as the
Planck scale is reached.

At this stage, we compare our results with the mixmaster universe in
the generalized uncertainty principle (GUP)~\cite{BM}. Considering a
close connection between $z=2$ Ho\v{r}ava-Lifshitz gravity and
GUP~\cite{Myungch}, there may exist a cosmological relation between
them. Fortunately, the chaotic behavior of the Bianchi IX model was
not tamed by GUP effects, which means that the deformed mixmaster
universe is still a chaotic system. This is mainly because two
physical degrees of anisotropy $\beta_{\pm}$ are considered as
deformed while the time variable is treated in the classical way.
This supports that our approach without quantization is correct.

Furthermore, it was shown that  adding $(^{4}R)^2$ (and possibly
other) curvature terms to the general relativity leads to the fact
that the chaotic behavior is absent~\cite{BC}. Hence it is very
curious to see why $(^{4}R)^2$ does suppress chaotic behavior, but
$-\frac{2}{\omega}(R_{ij}R^{ij}-\frac{3}{8}R^2)$ does not suppress
chaotic behavior.

In conclusion, the mixmaster universe has provided   another example
that $z=2$ Ho\v{r}ava-Lifshitz gravity has shown chaotic behavior,
as other chaotic dynamics of string or M-theory cosmology
models~\cite{DH}. This may be  because we did not quantize the
Ho\v{r}ava-Lifshitz gravity and we did study its classical aspects.

\begin{acknowledgments}
Y.S. Myung and Y.-W. Kim were supported by Basic Science Research
Program through the National Research Foundation (NRF) of Korea
funded by the Ministry of Education, Science and Technology
(2009-0086861). W.-S. Son and Y.-J. Park were  supported by the
Korea Science and Engineering Foundation (KOSEF) grant funded by the
Korea government (MEST) through WCU Program (No. R31-20002).
\end{acknowledgments}

\end{document}